\documentclass[5p,times]{elsarticle}

\usepackage[T1]{fontenc}

\usepackage[latin5]{inputenc}
\usepackage{times}
\usepackage[figuresright]{rotating}
\usepackage{float}
\usepackage{subcaption}
\usepackage{url}
\usepackage{setspace}
\usepackage{balance}

\usepackage{multicol}
\usepackage{multirow}
\usepackage{color}
\usepackage{graphicx}
\usepackage[implicit=false]{hyperref}
\usepackage{epsfig}
\usepackage{amsmath, amssymb}
\usepackage{comment}

\usepackage{capt-of}
\usepackage{multirow}
\usepackage{array}
\usepackage{caption} 
\usepackage{pslatex}
\usepackage{enumitem}

\usepackage{tcolorbox}
\usepackage{tabularx}
\tcbset{tab2/.style={colback=red!5!white,colframe=red!50!black,colbacktitle=red!40!white,
coltitle=black,center title}}

\usepackage{colortbl}

\usepackage{amssymb}

\journal{Computer Networks}

\begin{document}

\begin{frontmatter}



\title{Intelligent Zero Trust Architecture for 5G/6G Networks: Principles, Challenges, and the Role of Machine Learning in the context of O-RAN}


\author[inst1]{Keyvan Ramezanpour}

\affiliation[inst1]{organization={Marconi-Rosenblatt AI/ML Innovation Laboratory, \\ANDRO Computational Solutions, LLC},
            city={Rome},
            postcode={13440}, 
            state={NY},
            country={USA}}

\author[inst1]{Jithin Jagannath}

\begin{abstract}
In this position paper, we discuss the critical need for integrating zero trust (ZT) principles into next-generation communication networks (5G/6G). We highlight the challenges and introduce the concept of an intelligent zero trust architecture (i-ZTA) as a security framework in 5G/6G networks with untrusted components. While network virtualization, software-defined networking (SDN), and service-based architectures (SBA) are key enablers of 5G networks, operating in an untrusted environment has also become a key feature of the networks. Further, seamless connectivity to a high volume of devices has broadened the attack surface on information infrastructure. Network assurance in a dynamic untrusted environment calls for revolutionary architectures beyond existing static security frameworks. To the best of our knowledge, this is the first position paper that presents the architectural concept design of an i-ZTA upon which modern artificial intelligence (AI) algorithms can be developed to provide information security in untrusted networks. We introduce key ZT principles as real-time Monitoring of the security state of network assets, Evaluating the risk of individual access requests, and Deciding on access authorization using a dynamic trust algorithm, called MED components. To ensure ease of integration, the envisioned architecture adopts an SBA-based design, similar to the 3GPP specification of 5G networks, by leveraging the open radio access network (O-RAN) architecture with appropriate real-time engines and network interfaces for collecting necessary machine learning data. Therefore, this work provides novel research directions to design machine learning based components that contribute towards i-ZTA for the future 5G/6G networks.
\end{abstract}




\end{frontmatter}


\section{Introduction}\label{sec:introduction}

Wireless communication has become the key enabler of emerging technologies such as autonomous vehicles, vehicle-to-everything (V2X) networks, smart infrastructure, and internet-of-things (IoT)~\cite{akpakwu2017survey, shen2020internet}. The fifth-generation (5G) networks provide a massive volume of heterogeneous devices with seamless connectivity and computational resources for autonomous and intelligent operation \cite{jagannath2019machine, yao2019artificial}. Further, sixth-generation (6G) -- and beyond -- networks incorporate more agile radio environments, including satellite and unmanned aerial vehicle (UAV) communications, to provide a three-dimensional (3D) radio~\cite{jagannath2020redefining, liu20206g}. However, traditional network security frameworks have obvious weaknesses in providing security assurance in such a complex and dynamic network environment.

Traditional network security models assume a network perimeter, as the trust zone, which is protected against unauthorized access. Any subject operating in the trust zone, after appropriate authentication and authorization, is deemed trusted. However, due to the agile radio environment, mobility, and heterogeneity of next-generation tactical networks, identification of the network perimeter is challenging if not impossible. More importantly, such models allow lateral movement of subjects in the trust zone after authentication. 

The third generation partnership project (3GPP) has developed enhanced security frameworks specifically designed for 5G network architecture \cite{3GPPTS}. They introduce several security levels for various network functions including network access, user/application domains, and service-based architecture (SBA) security. The frameworks incorporate appropriate authentication, authorization and, key agreement protocols for the security of various technologies, such as device-to-device (D2D) and V2X communications, software-defined networking (SDN), and network function virtualization (NFV). 

Most existing security protocols assume a strong trust relationship among network entities and services providing authentication and authorization. Such assumptions can lead to serious security vulnerabilities. A few scenarios where these vulnerabilities are exploited to deploy privacy attacks, denial-of-service (DoS), man-in-the-middle, and impersonation attacks, are discussed in \cite{cao2019survey, Ramezanpour22Comnet}.

Zero trust architecture (ZTA) is a solution to address security requirements in a network with untrusted infrastructure \cite{NIST_ZTA}. 
A ZTA provides network assurance under the assumption that no subject, requesting access to the network resources, can be trusted even after initial authentication and authorization. Every access request is individually authorized and monitored during the access period for compliance with security policy rules. A dynamic trust evaluation for every access request is the key tenet of zero trust (ZT). The main function of a ZTA is authorizing the individual access requests by a subject rather than authorizing the subject requesting access.

The U.S. Department of Defense (DoD) introduces ZTA as a necessary paradigm shift in cybersecurity from the classical perimeter-based security to a data-centric model of security systems \cite{dod_zta}. In the perspective of this model, rather than a means for communication, networks are considered as a means for distributed data management; i.e., accessing, processing, transferring, and storing data through the network. Hence, the next generation security frameworks, which are based on ZTA, are expected to protect the data during this entire cycle. One of a few realizations of a ZTA-based security system for smart healthcare exploiting cloud services, through 5G networks, for data processing and management is recently introduced in \cite{chen2020security}.

Dynamic risk assessment and trust evaluation are key elements of a ZTA. We introduce the architecture of an intelligent ZTA (i-ZTA) which provides a framework to employ artificial intelligence (AI) engines for information security in untrusted networks. We also discuss the necessity for AI in obtaining a full realization of ZTA for next generation 5G networks. We discuss the adoptability of open radio access network (O-RAN) for the integration of such i-ZTA. Further, we argue how the multi-access edge computing (MEC) technology of 5G networks can be exploited to provide resource-constraint devices with the necessary computational resources for realizing the envisioned i-ZTA. 

The rest of the paper is organized as follows. First, in Section \ref{sec:impact}, we provide a brief impact statement to provide the readers the envisioned impact of this position paper. Section \ref{sec:whyZTA} introduces ZT principles and the necessity for the i-ZTA. Challenges of integrating i-ZTA into existing networks and the distinct features of next-generation networks enabling i-ZTA are discussed in Section \ref{sec:challenges}. The envisioned architecture and research directions for the i-ZTA is explained in Section \ref{sec:proposedZTA}, and the paper concludes in Section \ref{sec:conclusion}.

\section{Impact statement}\label{sec:impact}

Fifth-generation (5G) wireless networks are expected to handle a large volume of data generated by a wide range of devices including smartphones, autonomous vehicles, smart buildings, cities, and infrastructure. The members of society nowadays have their personal sensitive data stored on servers throughout the network and rely on the various access points to avail network-based services and applications. More importantly, the operation of future critical technologies such as autonomous vehicles, smart grids and emergency systems, that deal with public safety and national security, highly rely on network-based data management and processing. Zero trust architecture (ZTA) is an emerging and necessary framework for cybersecurity to provide the privacy and security of personal and sensitive data in such dynamic environment, which is also introduced by the U.S. Department of Defence (DoD) as the basis for next generation security systems. To the best of our knowledge, this is the first work that discusses how zero trust principles can become the driving force in ensuring security for future wireless networks (5G/6G). In addition, we introduce an O-RAN compliant conceptual intelligent ZTA framework based on artificial intelligence (AI) engines and discuss how the existing AI algorithms can be exploited to realize the premises of a ZTA. The ultimate goal of the paper is to open and accelerate a new research direction in AI for implementing the next generation of zero trust infused cybersecurity systems.

\section{Why Intelligent Zero Trust Architecture} \label{sec:whyZTA}

Ubiquitous connectivity through 5G networks is perceived by the U.S. DoD as a critical strategic technology that provides nations with long-term economic and military advantage \cite{dod_5G}. Next-generation networks are especially important for mission-critical communications and tactical edge networks (TEN), involving a large volume of heterogeneous and resource-constraint devices. They provide necessary computational resources (through cloud computing) and seamless, reliable, and robust connectivity through a wide range of new radio access technologies (RAT), including satellite, UAV, D2D, and massive beamforming communications. The 5G mobile networks are further expected to adopt a multi-RAT architecture in which different types of radio access (RATs) are unified to provide seamless connectivity \cite{galinina20155g, monteiro2018distributed}.

The deployment of a TEN, based on next-generation networks (5G/6G), is cost-effective (both in terms of CAPEX and OPEX) while programmable based on the needs of the particular environment. Hence, the deployment time of TEN with varying environmental needs to reduce significantly. However, perimeter-based security models exhibit weaknesses in providing network assurance in a heterogeneous and dynamic network environment. Further, the operation of intelligent TEN might heavily rely on cloud-based services for data management and processing. Hence, the data-centric model of ZTA for such highly mobile networks is necessary for information security.

Even if perimeter security frameworks provide carefully-tailored protocols for various functions of 5G networks, their static nature still allows lateral movements in the network perimeter. Either due to internal human errors, social engineering attacks, or dynamics of 5G networks, an authenticated subject (which is trusted) can acquire unauthorized access to sensitive resources. Hence, i-ZTA will be the key technology for secure communication and data sharing from core 5G to tactical edge networks.

Challenges of defining network perimeters, for enforcing security policies, and the potential lateral movements in traditional security architectures are the main propellants of ZTA \cite{teerakanok2021migrating}. As discussed in \cite{campbell2020beyond}, "\textit{trust is a vulnerability}" and the plausible solution to remove this vulnerability is a security architecture without any underlying trust assumption. The situation is exacerbated in 5G/6G architectures which employ NFV and SDN with the goal of providing programmability and scalability. In this network architecture, individual functions and services are deployed as virtualized software running on general purpose cloud infrastructure shared with untrusted third parties \cite{han2015network}. 

The rapid revolutionary developments in 5G network architecture aim at performance improvement while the security frameworks are lagging in addressing vulnerabilities in this new network environment. In virtualized network architecture and deployment, 1) network perimeter is diluted, and 2) network operates on untrusted infrastructure. However, security mechanisms in traditional cellular networks, especially authentication protocols, assume strong trust relationships \cite{cao2019survey}. The assumption of trust contradicts the emerging architecture and deployment model of beyond 5G networks. Hence, integrating ZTA with emerging virtual 5G/6G networks becomes a necessary security solution.

While recent research has focused on the necessity and advantages of ZTA for network security, deployment of a comprehensive architecture that controls the security process of all network accesses during its entire life-cycle is still an open problem \cite{buck2021never}. According to \cite{bertino2021zero}, ZTA provides a systematic approach to addressing challenges of traditional security frameworks, however, the deployment and management of ZTA is challenging. Existing proposals for ZTA demonstrate integration of partial zero-trust elements within a network security framework for addressing known security vulnerabilities and attack scenarios. In \cite{alevizos2022augmenting}, a distributed intrusion detection and prevention system (IDS/IPS), based on blockchain, is introduced to identify and remove compromised endpoints in a network. This work can be considered as an extension of classical IDS/IPS solutions to a distributed architecture consistent with ZTA solutions.

The problem of IDS/IPS within a ZTA architecture is also investigated in \cite{d2021building}. In this work, a solution to the problem of detecting security attacks is introduced by implementing zero-trust mechanisms at different layers of the open systems interconnection (OSI) model of the communication protocol stack. However, the promise of ZTA is not limited to detecting intrusion or security attacks. Rather, monitoring the entire life-cycle of a network session is a fundamental requirement of ZTA. In this regard, \cite{dimitrakos2020trust} proposes an authorization framework for IoT networks which integrates an attribute-based access control with a trust-level evaluation engine for continuous monitoring of a session during its entire life-cycle. Further, a trust evaluation mechanism, following a zero-trust strategy, is introduced in \cite{mehraj2020establishing}, for verifying the trustworthiness of services deployed on a cloud platform.

\subsection{Important of ZTA for Beyond 5G Networks}
According to the current research, intrusion detection and continuous verification of trust, between network entities, are the critical components of a ZTA. As discussed below, the DoD zero trust reference architecture expands the ZTA scope, in a systematic approach, by introducing seven ZT pillars that require ZT protection. In this paper, we introduce a conceptual ZTA framework that deploy the required ZT mechanisms using AI engines. We also propose a solution for integrating the proposed architecture with 5G O-RAN in an untrusted and virtualized network environment (potentially with cloud deployment). 

\begin{figure*}[t!]
	\centering
		\includegraphics[width=0.8\textwidth]{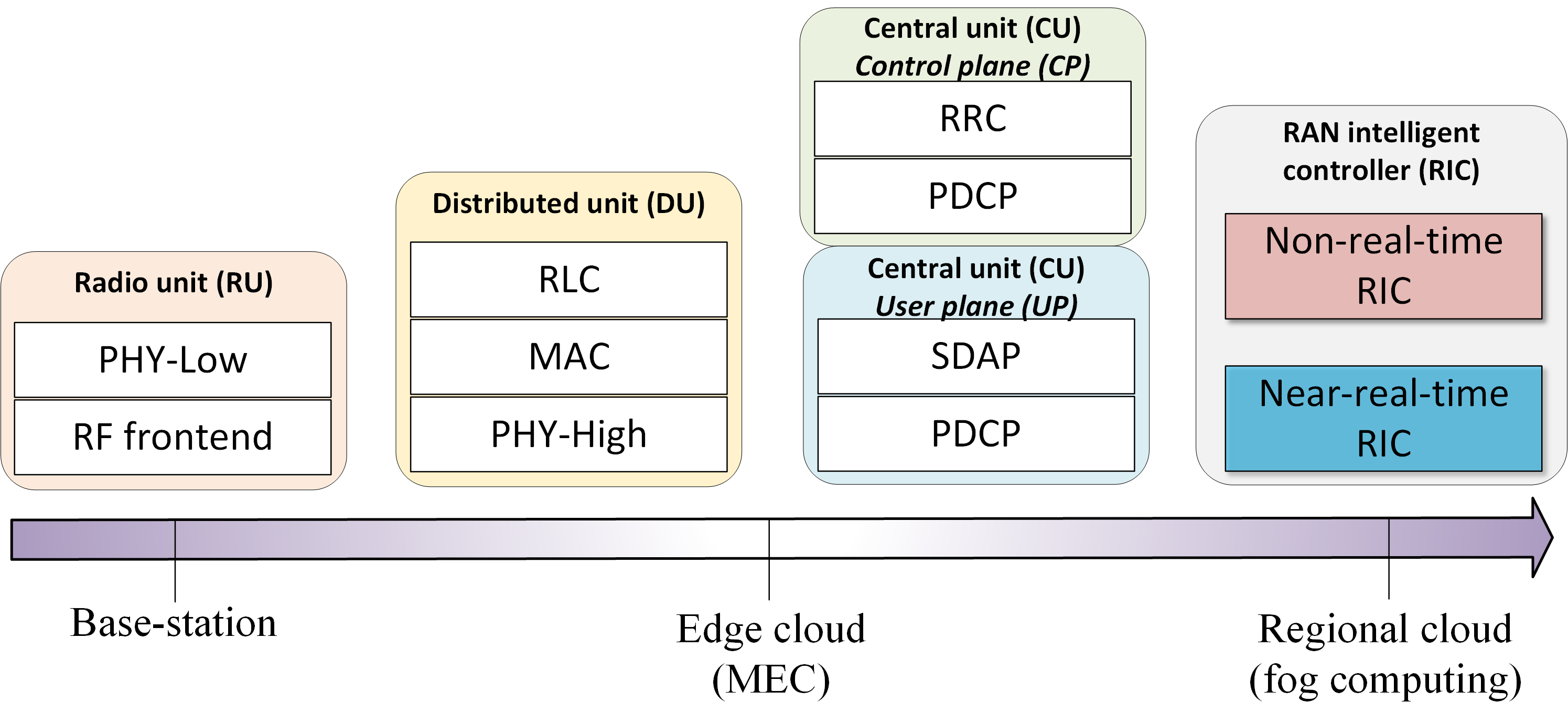}
	\caption{Disaggregated architecture of O-RAN protocol stack for distributed deployment of functional units on cloud platforms in a virtualized network.}
	\label{fig:ran_stack}
\end{figure*}

The O-RAN architecture introduces the concept of \textit{disaggregation} in which different layers of the communication protocol stack (in the RAN) are divided into separate function units as shown in Fig. \ref{fig:ran_stack}. A detailed description of this architecture is discussed in \cite{polese2022understanding}. Various function units of O-RAN can be deployed on different cloud platforms, from edge to fog computing, depending on the bandwidth and real-time processing requirements. In fact, this disaggregated architecture is the key feature of O-RAN in realization of NFV on cloud platforms which provides significant programmability and scalability advantages. The compromise is operating in an untrusted environment in which different network functions use shared third-party infrastructure to exchange information between different layers of the protocol stack. 

The O-RAN architecture also introduces RAN intelligent controller (RIC) that implements (closed-loop) control loops for network (RAN) optimization. The RIC is divided into two near-real-time (near-RT), with latency requirements between 10ms to 1s, and non-real-time (non-RT, with larger than 1s latency, components. The near-RT RIC hosts applications, called xApp, that manage QoS requirements of UEs. The non-RT RIC applications, called rApp, implement all management and optimization functions with a periodicity of larger than 1s including network orchestration, traffic routing and monitoring RAN components. The xApps can also include AI/ML engines that implement the decision engines of the near-RT RIC while rApps in the non-RT RIC control and manage the operation of the AI/ML engines.

The distributed architecture of O-RAN, and beyond 5G networks in general, with information exchange over open interfaces through Internet-based cloud platforms, broaden the surface of security attacks on beyond 5G networks. As discussed in \cite{polese2022understanding}, the open architecture of O-RAN can result in compromising the availability, data integrity, confidentiality, and AI/ML security attacks. Further, inclusion of third-party rApp and xApp in the RIC can provide attackers with the exploit of taking control of different nodes of the network. In this view, a ZTA framework, with real-time monitoring and risk assessment, and the objectives of maximum availability and least privileges, seems a necessary and promising solution to protect the beyond 5G infrastructure from emerging security threats.

Existing proposals for ZTA implementation focuses on IDS/IPS and continuous authentication mechanisms in the application layer, above the TCP/IP, in the Internet protocol stack. The main objective of these ZTA solutions is protecting databases from unauthorized accesses (detection and prevention at application layer). However, as discussed in the previous section, the DoD reference ZTA requires ZT controls over the entire access chain, from device and user to network environment and DAAS. Within this model, different layers of the O-RAN protocol stack in Fig. \ref{fig:ran_stack}, for establishing and managing the connection of mobile users to network resources, are the critical components of the network environment that require ZT controls. Due to the complexity and widespread adoption of 5G/6G networks for future critical applications, we believe that integrating a ZTA framework within the network architecture is crucial for protecting the infrastructure from security threats. The main objective of this paper is thus introducing such a comprehensive ZTA framework for integration within beyond 5G network architectures.

\subsection{Zero Trust Security Model} \label{sec:zta_sec_model}
In classical cybersecurity models, the authentication process establishes \textit{trust} for the network access control (NAC) in authorizing a user/device to access data, assets, application, services (DAAS). However, in a ZTA, a successful authentication alone does not imply trust, hence the name zero trust (ZT). Rather than a trust basis, the authentication is a prerequisite for access in the ZTA. The trust is evaluated using additional factors as discussed below.

\begin{figure}[t!]
	\centering
		\includegraphics[width=0.45\textwidth]{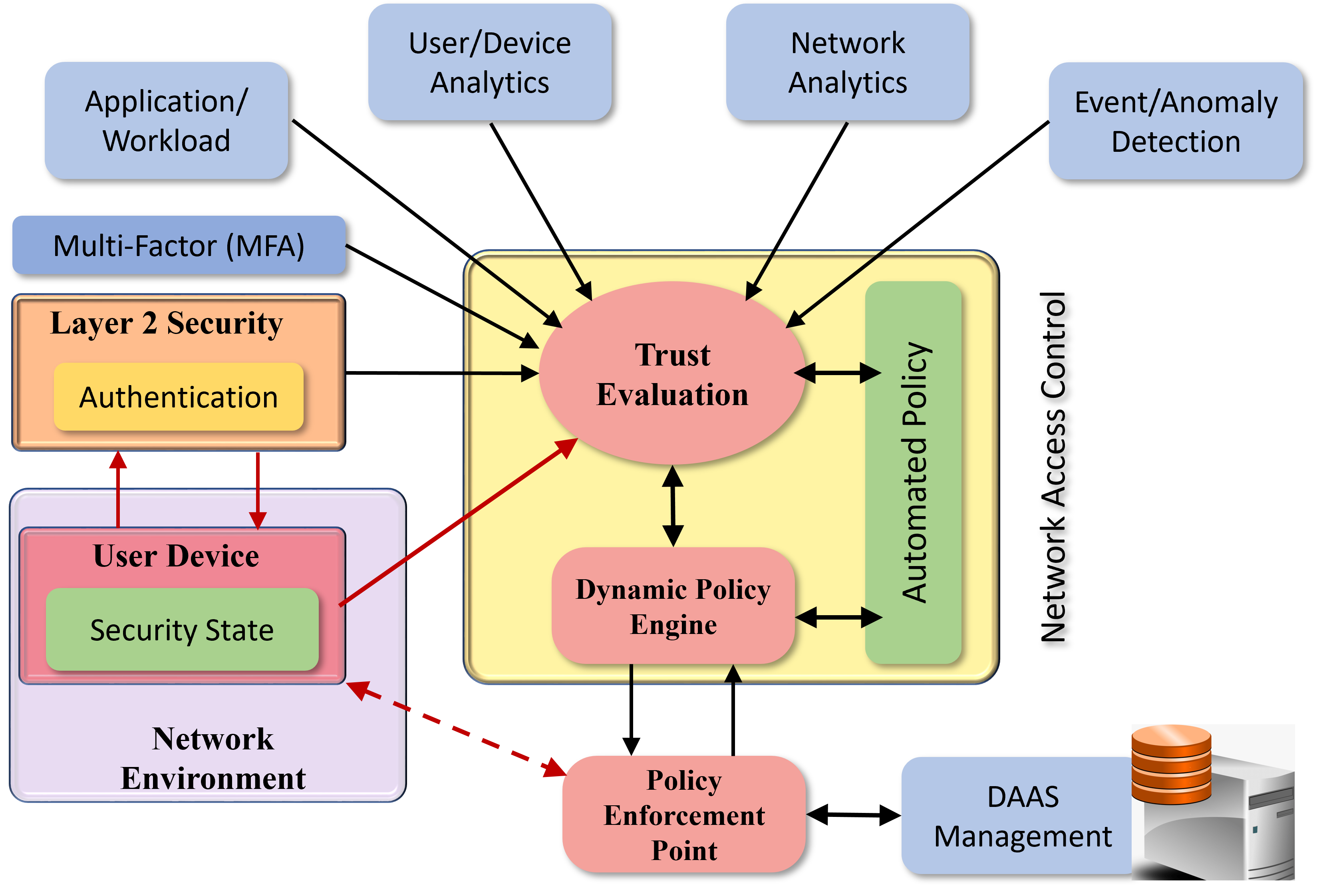}
	\caption{Basics of zero trust architecture (ZTA) for authentication and authorization including seven zero trust pillars.}
	\label{fig:basics}
\end{figure}

The U.S. DoD reference ZTA defines seven ZT pillars as: user, device, network/environment, application and workload, data, visibility and analytics, automation, and orchestration. Every pillar is a focus area in the implementation of a ZTA that requires appropriate ZT controls. The details of capabilities and requirements of the pillars is outlined in \cite{dod_zta}. The ZT engines in the seven pillars provide the necessary information for authorization. Under this model, as shown in Fig. \ref{fig:basics}, a trust evaluation engine becomes a critical component of a ZTA for access authorization. 

Since authentication does not imply trust, the NAC in a ZTA requires additional factors to evaluate trust and assess the risk of access by a user/device. In this model, the role of multi-factor authentication (MFA) becomes critical. Various MFA mechanisms (especially 2-factor) has become the mainstream in existing security systems, particularly using hardware/software tokens and bio-metrics for human subjects \cite{ometov2018multi, ometov2019challenges, jacomme2021extensive}. However, the DoD reference ZTA goes further and introduces continuous MFA (CMFA), a process which evaluates the authenticity of a user during an active session. The CMFA plays two critical roles in a ZTA: 1) supporting regular trust evaluation, and 2) providing a wider visibility on the user device and its network environment. 

In addition to the CMFA, the trust evaluation engine also requires event and anomaly detection as a a critical component for access authorization. An anomaly refers to suspicious activity in the network that might be malicious. An event is a change in the network, user and/or device behavioral patterns that might be malignant. The results of these engines help the NAC in implementing a fine-grained and dynamic policy for access authorization. The fine-grain access refers to micro-segmentation of the network in which different types of access requests (e.g., read or write to databases, accessing different applications and services) are authorized individually. The dynamic policy enables NAC to deny or grant accesses based on anomalies or events, detected in real-time, or requiring a device to pass extra CMFA processes in different network environments. Implementation of these engines are an important application of AI/machine learning which enables next-generation network security models.

The common features of the seven ZT pillars include real-time monitoring (user, device, network environment, visibility and analytics), risk assessment and trust evaluation (for access to data/application in an environment), dynamic policy and decision making (for granting access, automation, and orchestration). The i-ZTA proposed in this paper introduces the AI engines for realizing the main features of the seven pillars.

\subsection{Zero Trust Principles}
\begin{figure}[t!]
	\centering
		\includegraphics[width=0.4\textwidth]{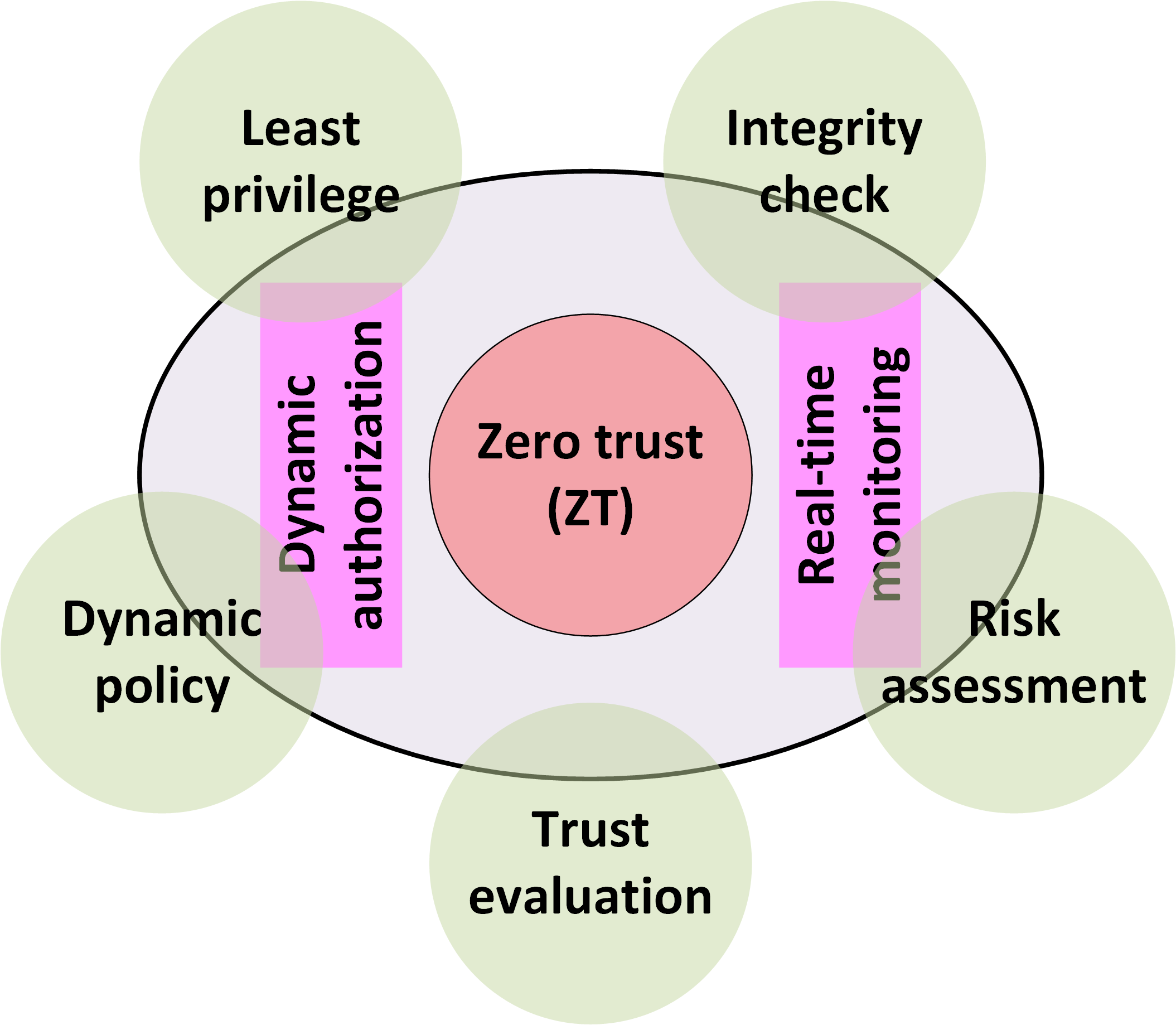}
	\caption{Key zero trust principles for information security in untrusted networks.}
	\label{fig:zt}
\end{figure}

The main tenets of zero trust (ZT) are outlined in the special publication 800-27 of the U.S. National Institute of Standards and Technology (NIST) \cite{NIST_ZTA}. The key ZT principles are summarized in Fig. \ref{fig:zt} and explained below.
\begin{itemize}
    \item \textbf{Zero Trust:} All network assets and functions, including devices, computing resources, and services, are considered untrusted irrespective of the location in the network. Hence, all communications must meet the same security requirements as third parties.
    \item \textbf{Trust/Risk Evaluation:} Trust evaluation and risk assessment are conducted for every access request. The assessment is carried out continuously (during the period of the access) and dynamically (based on situational conditions).
    \item \textbf{Least Privilege:} Any access, if granted, should be authorized with the least privileges. The access is only granted for a specific resource (depending on the sensitivity of the resource) and is not valid for a different resource. 
    \item \textbf{Dynamic Policy:} A dynamic policy is necessary for making the decision on granting access. The key decision factors include security state (credentials, software version/patches, location, etc.) and behavioral attributes of the subject and network assets. 
    \item \textbf{Integrity Check:} The security state of all network assets and requesting subjects are monitored continuously, preferably in real-time. The security posture of devices, and behavioral patterns of users/network assets, are evaluated with an automated system in terms of compliance with security policy rules.
\end{itemize}

\subsection{Intelligent MED}
The realization of ZT principles with static policies is overwhelming and challenging. \textit{Automated} real-time monitoring and \textit{dynamic} security evaluation are the key features of a ZTA. Further, with the growing volume of users, the ZTA components are dealing with \textit{big data}. Hence, intelligent monitoring, evaluation, decision-making (MED), using artificial intelligence (AI), are the critical enablers of ZTA in the next-generation networks.

\begin{figure*}[htbp]
	\centering
		\includegraphics[width=.99\textwidth]{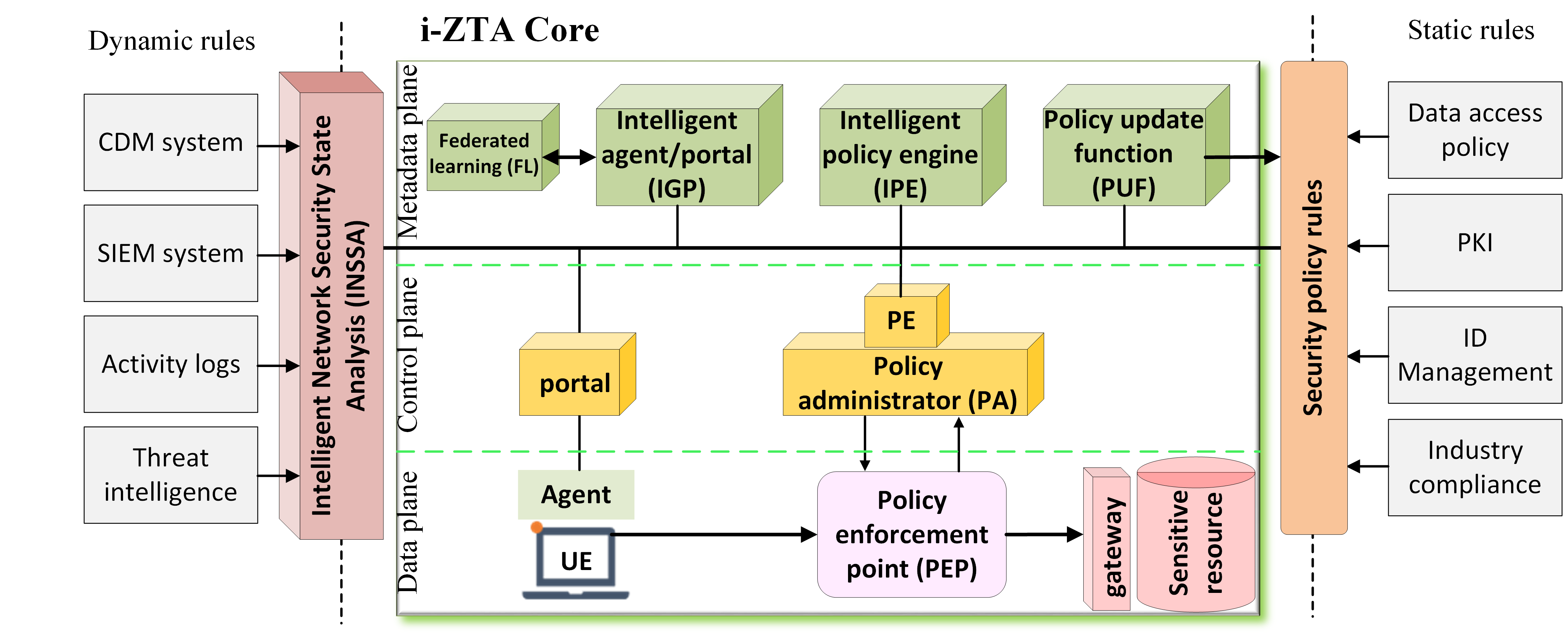}
	\caption{Logical components of the envisioned i-ZTA showing static security policy rules and AI engines for dynamic MED.}
	\label{fig:izta}
\end{figure*}

The MED components of an i-ZTA are shown in Fig. \ref{fig:izta}. The location of different blocks in this diagram reflects the logical interaction of the components and does not necessarily show their physical locations in the network. In this paper, we use the following terminology when referring to the i-ZTA. A \textit{subject} is any user, application, or service requesting access to a network resource. The network \textit{assets} refer to all devices, network infrastructure, and functions (including cloud services) involved in the communication. The network \textit{resource} contains sensitive information that must be protected against unauthorized access.

The core of an i-ZTA comprises of a policy enforcement point (PEP) and a policy decision point (PDP). The PEP is the first point of contact for access requests. It also establishes the connection between the subject and the requested resource if an access is granted. The decision on granting access is made by the PDP. It uses all available internal and external information about the security state of the subject and network assets for deciding. 

The information used by the i-ZTA core to grant and monitor a connection is provided by several peripheral modules as shown in Fig. \ref{fig:izta}. We divide these modules into two categories of static (right side of the figure) and dynamic (left side). The static modules (not specific to i-ZTA) include data access policy, public key infrastructure (PKI), identity (ID) management, and industry compliance. These modules, collectively, define the security policy rules for secure communication and integrity check rules. The policy rules can be dynamically adjusted by the i-ZTA core.

The dynamic modules in the diagram of Fig. \ref{fig:izta}, are the distinct features of an i-ZTA. These include continuous diagnostics and mitigation (CDM), threat intelligence (for identifying new security vulnerabilities), activity logs (behavioral information on user/assets and network traffic), and security information and event management (SIEM) for collecting information on long-term security state and potential attacks. 

In addition to the peripherals, the core PEP and PDP functions incorporate appropriate AI engines for realizing the entire MED chain. Intelligent agent and portal (IGP) is the AI engine of the PEP which provides devices with situational awareness. The processing engine of PDP is an intelligent policy engine (IPE) that makes decisions on granting access based on all information provided by INSSA, IGP, and policy rules. The details of the corresponding learning algorithms will be explained later in Section \ref{sec:proposedZTA}.

The i-ZTA of Fig. \ref{fig:izta} divides the network into three logical, and possibly physical, planes. Data communication between the subject and network resources is carried out in \textit{data plane}, which also includes the initial access request by the subject. The i-ZTA components (PEP and PDP) communicate in the \textit{control plane} for making decisions and configuring connections. These two planes also exist in the current 5G network architectures. The third plane of the i-ZTA is \textit{metadata plane} used for communicating all data required by the AI engines. 

\section{Challenges and Opportunities} \label{sec:challenges}

The realization of an i-ZTA, with real-time processing of big data might appear challenging. However, next-generation network architectures provide appropriate computational resources and interfaces for data collection needed by AI applications. In this section, we address part of the challenges and solutions for realizing the envisioned i-ZTA.

\subsection{Real-time Processing}
Seamless connectivity in beyond 5G networks implies that multi-RAT technologies are used dynamically in a single session of data communication \cite{chandrashekar20165g, ong2010optimal, yan2018smart}. Furthermore, heterogeneous devices, with different security specifications, credentials, privileges, and computing resources participate in the communication. Hence, real-time monitoring and security evaluation of all involving devices is the necessity of an i-ZTA.

Next-generation network architectures integrate NVF cloud computing for the realization of real-time functions for intelligent processing engines, with commercial off-the-shelf hardware. A promising example is the O-RAN architecture that provides (near-)real-time (10ms to 1s) RAN intelligent controller (RIC), central (CU), distributed (DU), and radio units (RU). While initial use cases of real-time engines in O-RAN focus on connectivity management in multi-RAT and quality of experience (QoE) optimization \cite{O-RAN}, integration of i-ZTA functions is an emerging critical application.

\subsection{Communication Overhead} \label{sec:comm_overhead}
Dynamic access in multi-RAT also implies a broadened attack surface. Third-party devices have now a wider range of access points to the network. They can manipulate an authorized user equipment (UE) to access network resources, or simply promote a hostile network environment to increase the perceived risk of access, hence, forcing the i-ZTA to decline access. 
Distributed location of network assets and massive volume of UEs has also facilitated the class of unintrusive precision cyber attacks which do not require privileged access to deploy the attack \cite{10.2307/26902668}. 

Detection and mitigation of such a broad attack surface require an analysis of network traffic by the i-ZTA, from UE to network assets. This analysis implies collection of big data in the network. The O-RAN architecture provides the E2 interface to collect data from CU and DU nodes for monitoring network assets. Also, O1/O2 interfaces are introduced for collecting machine learning data. Further, remote E2 and O1/O2 interfaces facilitate data collection from remote UEs and VNF. While the necessity of AI engines and interfaces in 5G architectures is mainly justified for network \textit{control} and \textit{optimization}, the deployment of i-ZTA is a third critical factor revealing the vital need for integration of AI to 5G networks.

\subsection{Computational Requirements}
The multi-RAT network and D2D communications allow attackers to exploit UEs without intrusion into the network. Hence, the envisioned i-ZTA requires all authorized devices to dynamically monitor their network environment for potential risks, as part of the intelligent MED. The major concern, however, is the limited computational resources of UEs, especially IoT node and sensor devices. We argue that multi-access edge computing (MEC) \cite{kekki2018mec} in 5G networks can be leveraged to address this issue.

The MEC is a prominent example demonstrating the unique capabilities of 5G networks, which brings the computing resources as close as possible to the network edge. Hence, a \textit{high volume} of devices, possibly with high \textit{mobility}, may have access to high performance \textit{computing} resources with \textit{low-latency connectivity}. The MEC adapts a similar service-based architecture (SBA) of 5G networks according to 3GPP specifications. The deployment of MEC can be considered as a mapping onto a network application function (AF) interacting with other network core functions.

\begin{figure}[t!]
	\centering
		\includegraphics[width=0.45\textwidth]{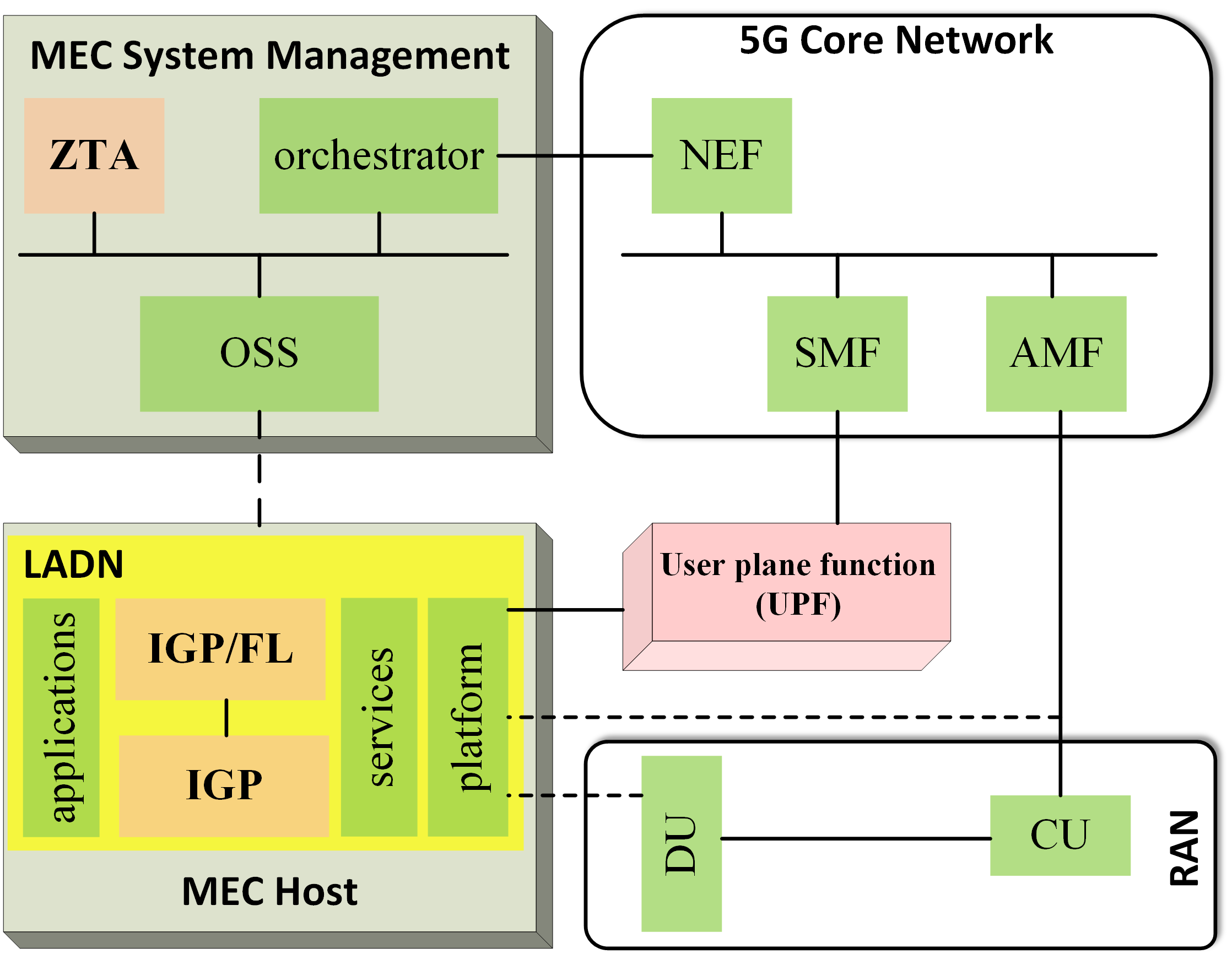}
	\caption{Architecture of multi-access edge computing (MEC) integrated at the edge of 5G network with i-ZTA core, IGP and federated learning (FL) components.}
	\label{fig:mec}
\end{figure}

An example of MEC architecture deployed in a local area data network (LADN) is shown in Fig. \ref{fig:mec}. The MEC orchestrator is a 5G AF with centralized functions for managing the operation of MEC hosts. It also interfaces with the network exposure function (NEF) in the 5G core for overall management. A unique feature of 5G networks enabling the integration of this MEC architecture is the exposure of the network core to the LADN. The 5G core network is able to steer traffic to the applications in the LADN, where the MEC host operates.

The MEC hosts are deployed at the edge of 5G RAN to minimize latency and improve user QoE. An interesting feature of this architecture is the exposure of MEC hosts to radio information provided by the CUs and DUs of the RAN. 
The MEC platform may use radio information, e.g., signal power and quality, to further reduce latency by avoiding unnecessary routing traffic via the core network. 

While MEC enables the deployment of some intelligent MED components, it is also protected by the i-ZTA. 
The 5G core incorporates appropriate functions such as NEF and policy control (for traffic steering) and unified data management (for user authentication, authorization, and service continuity) to provide untrusted AFs with requested services. 
While these functions provide static security measures, the i-ZTA core (Fig. \ref{fig:izta}) provides the dynamic security measures to authorize access to the MEC and monitor the sessions.

\section{Envisioned Intelligent Zero Trust Architecture and Research Directions} \label{sec:proposedZTA}
In this section, we introduce a novel unified framework (i-ZTA) with AI engines for information security in untrusted networks. The envisioned i-ZTA opens new research areas in the application domain of AI for information security. The key elements of the i-ZTA include:
\begin{itemize}
    \item \textbf{IPE (Intelligent Policy Engine):} It employs an AI \textit{trust algorithm} to authorize access requests based on subject privileges and security state, security policy rules, the network state, and a score that reflects the confidence level of the access. Specifically, the IPE is envisioned to employ reinforcement learning (RL) to maximize usability with the least privileges.
    
    \item \textbf{INSSA (Intelligent Network Security State Analysis):} This engine can employ such models as a graph neural network (GNN) for the security state of the network. It carries out the risk assessment in accessing a given resource in the network. The INSSA also implements an anomaly detection to identify potential attacks.
    
    \item \textbf{IGP (Intelligent aGent/Portal):} It is the user AI engine to model the security state of a subject. The IGP analyzes the security posture of the network traffic to the subject and provides it with environmental awareness. The learning objective of IGP is to keep a high confidence level of the subject in accessing the network resources.
\end{itemize}

\subsection{Overall Architecture}
\begin{figure*}[htbp]
	\centering
		\includegraphics[width=.99\textwidth]{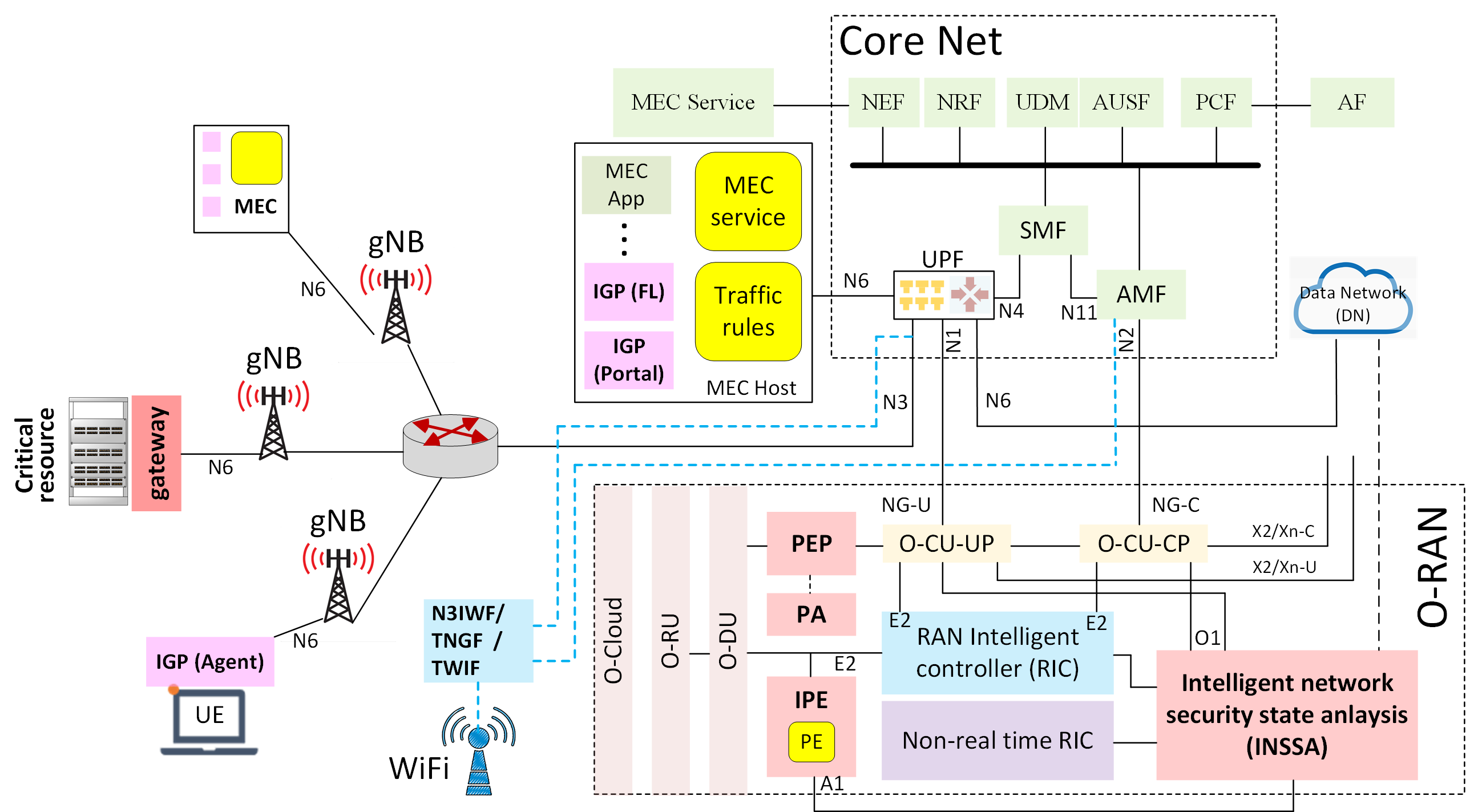}
	\caption{Architectural design of a 5G network, leveraging O-RAN, with integrated i-ZTA framework.}
	\label{fig:5gzta}
\end{figure*}

The architecture of the i-ZTA integrated in the O-RAN architecture is shown in Fig. \ref{fig:5gzta}. It exploits the O-RAN real-time processing and data collection for realizing the intelligent policy engine and AI network state analysis, and the MEC for intelligent agent and portal components of MED. The O-RAN further provides appropriate interfaces for near-real time monitoring of remote devices and services.

In our envisioned i-ZTA, the PEP is divided into three components: agent, portal, and gateway. The agent is a lightweight software module on every network asset requiring access to the resources. The portal, residing on the PEP, performs a similar task but is intended for resource-constraint devices, such as IoT and sensor devices. 
The gateway is an agent residing in front of the network resource and is directly configured by the policy administrator (PA). 

The agent and the portal incorporate AI algorithms, that work together in a federated learning approach, as described below. We refer to the learning components of the agent and the portal as IGP, collectively.

\subsection{Intelligent agent/portal (IGP)}
We define the environmental awareness (ENA) of a subject as the first tenet of trust evaluation. The role of IGP is to provide the network assets, that require access to the resources, with an \textit{ENA score}, and a model for their security posture. The subjects with higher ENA values might obtain higher confidence scores in risk assessment by the IPE. 

We envisage the agent to employ a reinforcement learning (RL) engine that conducts three main tasks: 1) It analyzes the traffic of the device in the network which provides an initial risk assessment on the network environment; 2) It learns the flow of unnecessary communication in the device that may reduce its confidence level in accessing certain resources; 3) It provides a model for the communication pattern of the device which is passed along with an access request to the PEP for overall risk assessment. As discussed in Section \ref{sec:zta_sec_model}, dynamic authorization is an important feature of a ZTA framework (Fig. \ref{fig:basics}). Reinforcement learning enables i-ZTA to learn from observations in real-time (user/device behaviors, events in accessing network resources and newly detected anomalies) and adjust trust policies accordingly. 

The portal is divided into two major components: 1) access request management from resource-constraint devices without computational capability to host the intelligent agent, and 2) a learning component that supports federated learning of the agents. We envision the importance of federated learning for collaborative and distributed learning of a comprehensive model for the network environment. 

As discussed above, every agent employs RL to learn its network environment and best practices in communicating with network entities. The RL model used by multiple agents can be a common model, trained in the federated learning approach with the following clear advantages.
\begin{itemize}
    \item By aggregating the experience of multiple agents, a more comprehensive model of the local network environment is trained by distributed subjects.
    \item The \textit{visibility} of individual agents on the network environment is increased.
    \item It provides subjects with a model for a network environment that is able to detect distributed attackers exploiting multiple subjects. 
\end{itemize}

\subsection{Intelligent network security state analysis (INSSA)}
The second tenet of trust evaluation is the dynamic risk assessment for every access request. In a 5G network, the mobility of heterogeneous devices in a varying environment calls for a dynamic model of the network state that provides information about the risks of accessing a particular resource by a given subject. We propose using a graph neural network (GNN) to model the state of the 5G network which is of particular interest in risk assessment by the IPE.

Graph neural networks have been shown to be successful as a scalable approach for resource allocation in large area wireless networks \cite{ruiz2021graph}. In these applications, the GNN models the channel state between pairs of communicating nodes and the goal is an optimal allocation of spectrum resources to the nodes. Further, recurrent GNNs have been shown successful in space-time modeling of data \cite{nicolicioiu2019recurrent}. The INSSA employs a recurrent GNN to model the communication patterns of a 5G network, over space and time, and the goal is to assign risk scores (R-Scores) to the nodes such that the overall metric of security assurance in granting an access is maximized. 

The envisaged INSSA employs reinforcement learning to meet the following objectives: 1) compliance with a set of security policy rules, 2) authorizing accesses with the least privileges, and 3) maximizing network usability. 
The RL algorithm dynamically assigns appropriate scores to all nodes in the network so as an \textit{assurance score} (as the reward) is maximized, by inspecting how strictly the policy rules are met by the nodes while the network is available to all users. The training process is accelerated by transmitting periodic test vectors to the devices and network assets. The response of the assets to the tests are used in an iteration of the training.

A second critical task of the INSSA is anomaly detection. The goal of this task is to detect and prevent potential attacks, such as DoS and distributed DoS (DDoS), that target the PEP. Additionally, INSSA protects the network against a more subtle DoS attack which we call intelligent DoS (IDoS) in this paper. As discussed above, the i-ZTA incorporates environmental information of the network to evaluate the access request. A potential attacker can indirectly make the PE deny access by promoting a hostile network environment, hence increasing the risk of granting the access. The INSSA uses the GNN model of the network to detect such activities and potential attacks.

\begin{figure*}[t!]
	\centering
		\includegraphics[width=0.6\textwidth]{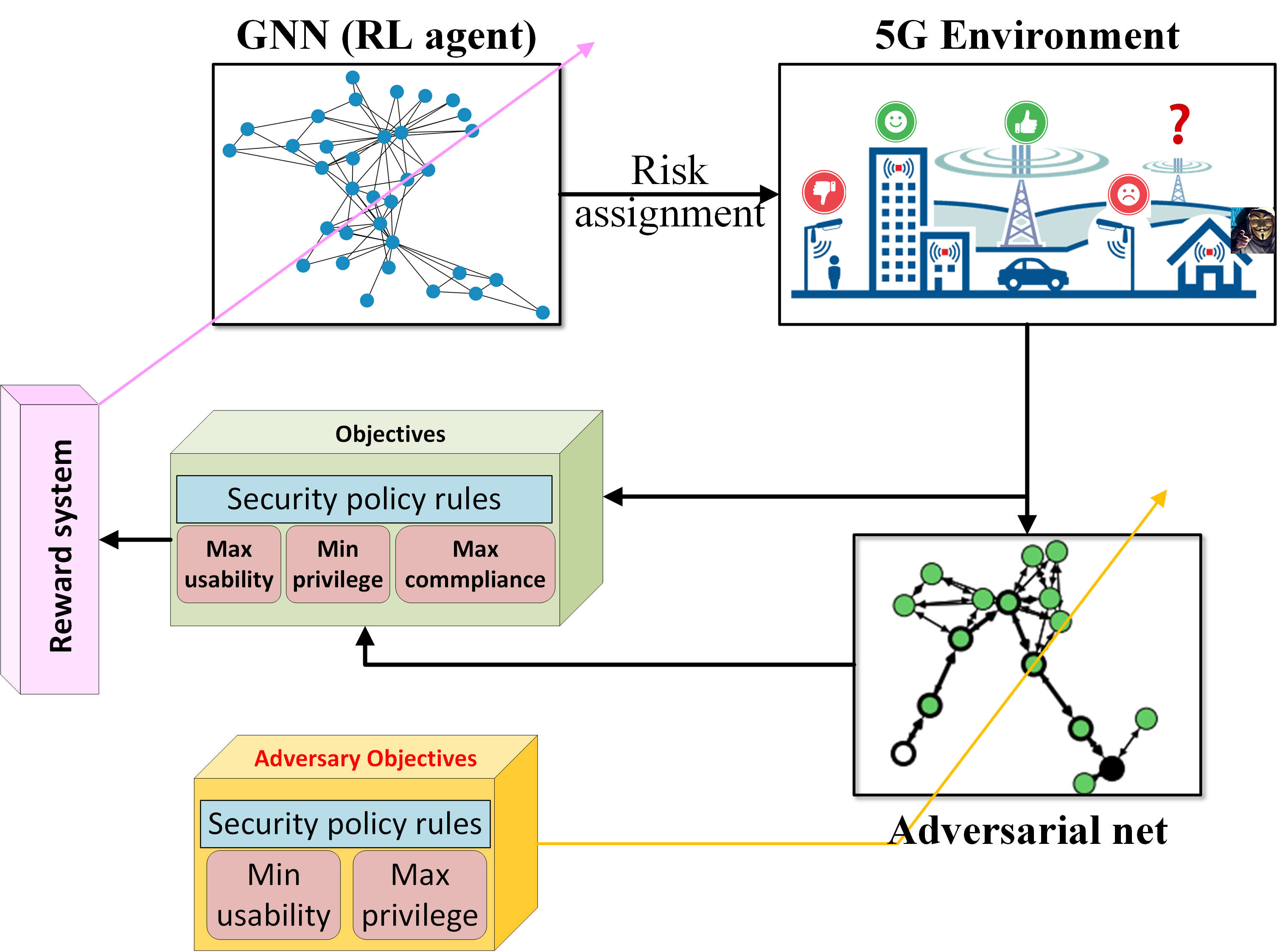}
	\caption{Adversarial learning methodology to maximize network assurance with objectives of minimum \textit{privileges}, maximum \textit{usability} and strict \textit{compliance} with security policy. }
	\label{fig:adversarial}
\end{figure*}

The INSSA follows an adversarial learning approach \cite{chen2020survey} for risk assessment and anomaly detection. In addition to the node risk scores, the risk of operating in a network environment contributes to the overall network assurance score. The block diagram of Fig. \ref{fig:adversarial} illustrates the flow of the proposed adversarial learning. While the GNN (as the RL agent) attempts in maximizing network usability with the least privileges, an adversarial network is trained with the objective of maximizing privileges compliant with the security policy rules. The result is a GNN model of the network trained to assess the risk of access in the presence of intelligent distributed attackers exploiting network assets.

The integration of risk assignment and anomaly detection into a single learning model is beneficial in the view of \textit{meta reinforcement learning} \cite{wang2016learning, duan2016rl}. Intrusion detection systems (IDS) using deep learning has been a popular and classical example of AI application in the field of cybersecurity for detecting malicious behavior \cite{vinayakumar2019deep, sultana2019survey, zhou2020building}. However, these IDS systems rely on diverse and heterogeneous malware datasets and can still be vulnerable to zero-day attacks (not observed in training). Alternatively, the few-shot learning perspective to meta-learning and its capability in learning new tasks quickly is essentially important in anomaly detection. The recurrent GNN model of the INSSA is also beneficial as meta-learning algorithms require a memory of the last action and states which is implicit in recurrent neural networks.

\subsection{Intelligent policy engine (IPE)}
The endpoint of trust evaluation is the IPE. Like the agent/portal and INSSA, the IPE incorporates an AI engine to make the final decision on granting a requested access based on the agent and network state. The IPE employs a neural network with long- and short-term memories to evaluate the risk of granting access to an agent based on all previous activities of the agent and the network. The IPE provides a C-score as the confidence-level of the access.

The IPE policy is optimized through an RL algorithm to minimize the probabilities of false positives and false negatives. After making a decision (access grant or denial), the IPE monitors the security state of the session (how strictly the agent conforms to the security policy rules). 
The IPE also receives the future state of the agent from the INSSA to evaluate the reward return corresponding to the decision. The IPE uses the collected information to evaluate the risk of the agent for its future transactions.

The memory of the IPE policy is an important feature for risk assessment. A potential intelligent attacker does not deviate from the security policy rules with observable traces. Rather, it attempts in exploiting multiple network assets by taking incremental steps toward malicious activities or unauthorized access to sensitive information, distributed over space and time. While the spatial security state is modeled by the INSSA, the IPE provides the temporal model.

It is important for the IPE to incorporate all previous and future states of the subject and the network for risk assessment. We divide the learning policy of the IPE into two sub-components with long- and short-term memories. The short-term memory allows granting access to agents which corrected their security state over time. The long-term memory enables IPE to detect adversaries exploiting vulnerabilities with incremental steps over time. 

\begin{figure}[t!]
	\centering
		\includegraphics[width=0.45\textwidth]{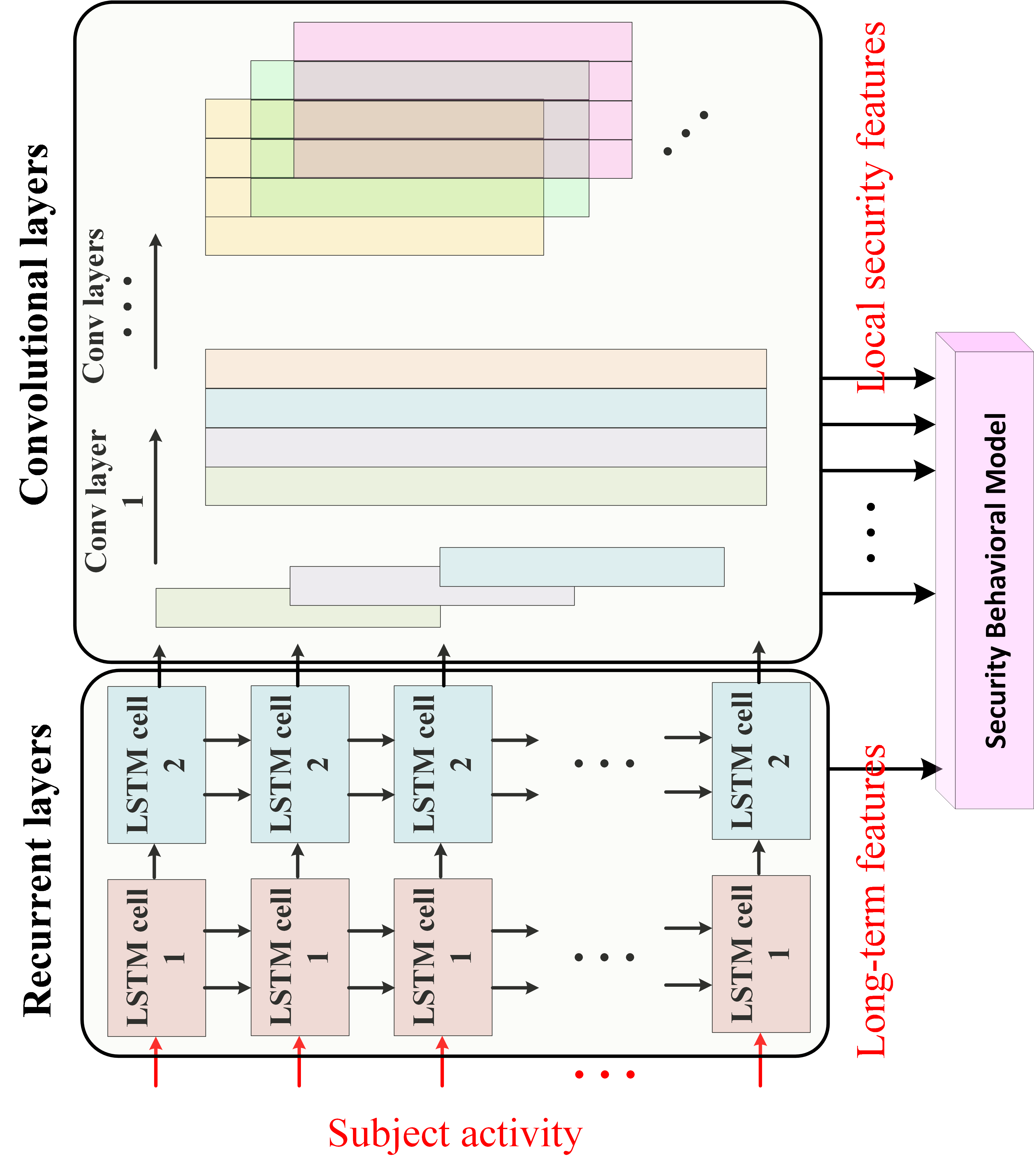}
	\vspace{-0.2cm}
	\caption{Example of Neural network of the intelligent policy engine (IPE) extracting medium- to long-term information (recurrent layers) and local security features (Conv. layers) of network activitis. }
	\label{fig:ipe}
\end{figure}

An example concept-level architecture of the IPE neural network (NN) with long- and short-term information is shown in Fig. \ref{fig:ipe}. The NN consists of a few recurrent layers followed by convolutional layers. It has been demonstrated that recurrent NN, such as long short-term memory (LSTM), is powerful in temporal modeling of a time-series signal and convolutional layers are capable of filtering noisy spectral components and extracting local features of the signal \cite{sainath2015convolutional}. Hence, the IPE NN extracts temporal information on the security behavior of the subject, over medium to long time periods, while the convolutional layers extract local (shorter time periods) security features of the subject. 

It is envisaged that all components and AI engines, introduced in this section, work together in a cohesive framework to meet the enhanced security needs of military as well as commercial 5G/6G networks in the future. Using this framework, sensitive applications may benefit from widespread adoptibality and low-cost deployment of these networks without compromising information security. 

The summary of acronyms used in the paper has been provided in Table \ref{tab:AcroZTA} and Table \ref{tab:Acro}.

\begin{table}[h!]
    \centering
    \caption{Acronyms Specific to i-ZTA \label{tab:AcroZTA}}
    \begin{tcolorbox}[tab2,tabularx={|p{1.5 cm}||p{9 cm}}]
      \textbf{Acronyms}   & \textbf{Definitions}  \\ \hline\hline
      CDM   & Continuous Diagnostics and Mitigation \\ \hline
      CMFA  & Continuous Multi-Factor Authentication \\ \hline
      DAAS  & Data, Asset, Application, Service \\ \hline
      DDoS  & Distributed Denial-of-Service \\ \hline
      IDoS  & Intelligent DoS \\ \hline
      ENA   & ENvironmental Awareness \\ \hline
      FL    & Federated Learning \\ \hline
      GNN   & Graph Neural Network \\ \hline
      IGP   & Intelligent aGent and Portal \\ \hline
      INSSA & Intelligent Network Security State Analysis \\ \hline
      IPE   & Intelligent Policy Engine   \\ \hline
      MED   & Monitoring, Evaluation, Decision making \\ \hline
      NAC   & Network Access Control \\ \hline
      PA    & Policy Administrator \\ \hline
      PDP   & Policy Decision Point \\ \hline
      PEP   & Policy Enforcement Point \\ \hline
      RL    & Reinforcement Learning \\ \hline
      SIEM  & Security Information and Event Management \\ \hline
      ZTA   & Zero Trust Architecture \\ \hline
   
    \end{tcolorbox}{}
\end{table}

\begin{table}[h!]
    \centering
    \caption{General Acronyms (5G network) \label{tab:Acro}}
    \begin{tcolorbox}[tab2,tabularx={|p{2 cm}||p{10 cm}}]
      \textbf{Acronyms}   & \textbf{Definitions}  \\ \hline\hline
      AF    & Application Function \\ \hline
      CU    & Central Unit \\ \hline
      DU    & Distributed Unit \\ \hline
      LADN  & Local Area Data Network \\ \hline
      MEC   & Multi-access Edge Computing \\ \hline
      N3IWF & Non-3GPP InternetWorking Function \\ \hline
      NFV   & Network Function Virtualization \\ \hline
      O-RAN & Open Radio Access Network \\ \hline
      QoE   & Quality of Experience \\ \hline
      RAT   & Radio Access Technology \\ \hline
      RIC   & RAN Intelligent Controller \\ \hline
      SBA   & Service-Based Architecture   \\ \hline
      SDN   & Software-Defined Networking \\ \hline
      TEN   & Tactical Edge Network \\ \hline
      TNGF  & Trusted Non-3GPP Gateway Function \\ \hline
      TWIF  & Trusted WLAN InternetWorking Function \\ \hline
      UE    & User Equipment \\ \hline
      VNF   & Virtual Network Function \\
   
    \end{tcolorbox}{}
\end{table}

\subsection{Realizing i-ZTA within O-RAN RIC}
The architecture of O-RAN with xApps and rApps of the RIC and access to open interfaces for collecting data, monitoring and controlling RAN components provides a highly flexible and programmable framework for deployment of different control engines including network access control (authorization) of the i-ZTA. In this section, we briefly discuss a realization scenario of deploying i-ZTA within the O-RAN RIC through xApp and rApps. A complete deployment leveraging cloud platforms, including MEC and fog computing, for training, verification and execution of AI/ML engines is a subject of future research. For a detailed description of different open interfaces in the O-RAN architecture, we refer an interested reader to~\cite{polese2022understanding}.

Among the three main i-ZTA engines, IPE is the heart of authorization with near-realtime (RT) operation. This engine evaluates trust and makes decisions on granting/denying for every access request to a network resource. The proposed IPE consists of three components: 1) environment visibility and analytics (EVA), 2) user behavioral model (UBM), and 3) trust evaluation engine (TEE). Every component is implemented as a microservice in the near-RT RAN intelligent controller (RIC). Hence, the components of IPE are implemented as xApps within the O-RAN architecture. In short, EVA and UBM provide information about the radio environment and user network usage, respectively. The TEE uses these information to evaluate the trust for granting access to a resource. The decision is made based on a least privilege policy, user priority and the sensitivity of the requested resource.

The EVA component of the IPE collects required data over the E2 interface of the O-RAN. The xApp implementing EVA subscribes to the distributed units (DU) of the O-RAN to receive key performance indicators (KPI) on the PHY, MAC and radio link control (RLC) for specific UEs. Important metrics collected by EVA include information on channel state and beamforming (location information of UEs), resource utilization (user activity) and link reliability (potential interference). These metrics provide trust engines with a fine-grained visibility on the radio environment for every UE.

The UBM xApp in the IPE subscribes to the central units (CU) of the O-RAN, via E2 interface, to collect information mainly on the traffic patterns. It also uses the KPIs pertinent to resource utilization, throughput, and load of the gNBs. The data collected for the gNB provides information on the number of users, interference level, and the presence of potential adversaries and jammers. The quality of service (QoS) metrics of the traffic flows, network load and throughput can be used as indicators of interference and jamming environment but also as metrics for prioritizing UE accesses. The UBM implements AI-based modeling of UE and network traffic patterns. The learnt pattern is also used for event detection. We point out that an event is not necessarily malicious. As an example, connection of a new UE to the network or handover of existing UEs are malignant events. However, an unexpected and abrupt change in a user traffic or QoS can be suspicious. The trust engine uses the traffic patterns and detected events in granting accesses.

The TEE in the IPE implements the trust evaluation engine. It uses the information and behavioral models provided by the EVA and UBM xApps as part of the features used in decision making. The TEE also collects data about UEs from IGP via the A1 interface. (As explained below, the IGP is implemented in the non-RT RIC.) The IGP provides TEE with information about user activity in the global network domain including Internet traffic and connections over non-3GPP access (e.g., 5G connection through WiFi access). This information complements those provided by EVA and UBM on the user activity over the 3GPP access (gNB) and radio environment. Based on these information, the TEE outputs a soft-decision metric, e.g., a real number in the range 0 (min trust) to 1 (max trust), which also depends on the sensitivity of the requested resource. It further generates a hard decision metric on grant/deny for the access. The soft-decision metric can be used in prioritizing accesses in congested conditions with limited physical resource blocks.

The second main component of the i-ZTA is IGP that provides users with environmental awareness. As discussed above, IGP is implemented as an rApp in the non-RT RIC within the Service Management and Orchestration (SMO) framework of the O-RAN. Three main considerations for deploying the IGP in the non-RT domain follows. First, the IGP collects information on the users traffic in the global network including data network and non-3GPP access. This information is available in the SMO which hosts the non-RT RIC. Second, monitoring network traffic does not require fine time resolution as in near-RT RIC (between 10ms and 1s). The near-RT resolution is more suitable for packet-level timing and inspection. Third, non-RT operation of the rApp implementing the IGP implies lower communication overhead and online computational requirements for this engine. Hence, the IGP can be hosted on lower complexity computing platforms equipped with general purpose processors.

The IGP rApp subscribes to CU of the O-RAN over O1 interface (and O2 interface when O-RAN components are deployed on the O-Cloud) to collect data. It also interfaces with RU and DU (over O1/O2) mainly for network monitoring and CMFA as described later. The data consumed by IGP includes user session management information such as number of sessions, traffic statistics, data network traffic, handover events and user mobility. The CU interfaces with User Plane Function (UPF) and the Access and Mobility Management Function (AMF), via N3 and N2 interfaces, respectively. Further, the internetworking functions of 5G (responsible for establishing connections through non-3GPP access networks) interface with UPF and AMF over N3/N2 interfaces. Hence, the CU can provide IGP with required information about users network activity. As discussed above, the IGP also sends this information, in a compact representation (enrichment information), to the TEE over A1 interface for trust evaluation.

The second task of IGP is to provide UEs with a mechanism for network monitoring. The information about user network activity (sent over A1 interface) helps TEE to evaluate the trust of accesses requested by UEs. The IGP is also responsible for providing UEs with a mutual trust evaluation on network environment. For this task, the IGP collects KPI reports from O-RAN components (RU, DU and CU) over O1 interface. The KPIs used for this task include registration success rate of slices, integrity KPIs (uplink/downlink latency), slice throughput, upstream/downstream N3 interface and UE throughput. By learning the temporal pattern of the performance metrics, the IGP derives a behavioral model of the network entities (hardware and software components) over time. Based on this model, the IGP provides UEs with an assurance metric (e.g., a real scale between 0 and 1) indicates trustworthiness of a traffic flow (bearer) to a UE.

Similar to IPE and IGP, the INSSA engine collects data from the CU of the O-RAN corresponding to all UEs. The INSSA is responsible for constructing a global model of the network. The objective of the INSSA is twofold: 1) assigning risk scores to UEs and network devices (RAN components), 2) global anomaly detection. The INSSA can be deployed on a regional cloud (e.g., fog computing) with data collected over O1 interface.

The above i-ZTA functions provide soft-decision mechanisms for verifying the trust of network components using a rich amount of data available through the O-RAN interfaces. In addition to these mechanisms, we also desire a trust verification mechanism with cryptographic proofs. For this purpose, the i-ZTA also includes a CMFA engine implemented as a rApp in the non-RT RIC of the O-RAN. This engine evaluates trust, in a continuous process, at two levels: 1) trust between i-ZTA functions, and 2) trust between i-ZTA and O-RAN components. 

The CMFA engine can employ next hop chaining counter (NCC) parameters along with cryptographic Hash functions to verify liveliness and integrity of different network components and functions. The CMFA engine triggers authentication requests for every component. Each network component and i-ZTA function also maintains a local NCC parameter. Upon receiving an authentication request from the CMFA engine, an i-ZTA function (xApps corresponding to IPE and rApps implementing IGP) responds back with the HD5 Hash of its software image and its local NCC. In the authentication request, the CMFA engine also sends the Hash of its own software image with the local NCC of the target function. (The CMFA engine sends the authentication requests over O1 interface to the xApps.) Hence, the target i-ZTA function can also verify the trustworthiness of the requesting CMFA engine. This process establishes the trust between i-ZTA functions.

For verifying the trustworthiness of O-RAN components, the CMFA engine can use any of the IPE xApps or IGP rApps as a bridge to forward authentication requests. We point out that all O-RAN components (including xApps) support X.509 certificates for authentication. When an IGP rApp receives an authentication request for O-RAN components (RU, CU or DU), it forwards the request within an ZT-AUTH message over O1 interface to the target component. This component responds back with the Hash of a shared secret (established during the initial authentication) and its local NCC. (The component also updates its local NCC upon receiving every ZT-AUTH message.) Similarly, the IGP sends authentication requests to IPE xApps through O1 interface for O-RAN components. The xApps then forward the request within an ZT-AUTH over E2 interface to the target component. The rApps and xApps receive the response from O-RAN components, integrate them within the Hash of their own image and local NCC, and forward the entire response to the CMFA engine for verification.

\section{Conclusion} \label{sec:conclusion}

Network assurance in the untrusted environment of 5G/6G networks demands dynamic authorization, risk assessment, and monitoring of network assets. Realization of zero trust (ZT) principles, necessary for providing information security in such environments, require real-time processing of big data. In our opinion, the envisioned intelligent architecture (i-ZTA) in this paper can help enforce ZT principles for tactical and commercial application by leveraging distinct technologies of 5G networks as the key enablers of the i-ZTA. While instances of ZTA realizations for smart IoT networks (relying on cloud-based services for data management and processing) are emerging, we discussed how AI engines of the i-ZTA facilitate handling big data encountered in a typical ZTA framework in the context of 5G/6G.

The i-ZTA adopts an SBA-based design with AI engines for realizing ZT principles in untrusted networks. The i-ZTA core includes the intelligent policy engine (IPE) and the intelligent agent portal (IGP) for dynamic authorization of access requests. The former uses reinforcement learning, with the objective of maximizing an assurance score, and the latter uses federated learning to provide users with environmental awareness score (EVA). Dynamic monitoring of the network assets is also realized with the intelligent network security state analysis (INSSA) which employs graph neural networks (GNN) for network modeling and adversarial learning for risk assessment. It is our sincere hope that the i-ZTA vision for 5G/6G discussed in this paper motivates new research directions in the application of AI for providing information security in untrusted networks. 


 \bibliographystyle{elsarticle-num} 
 \bibliography{Andro1}

\subsection*{  }
\noindent \textbf{Keyvan  Ramezanpour} completed a PhD in Electrical Engineering in the Bradley Department of Electrical and Computer Engineering at Virginia Tech. Dr. Ramezanpour currently is a senior researcher at the Marconi-Rosenblatt AI/ML Innovation Lab at ANDRO Computational Solutions. His research expertise include secure implementation of cryptographic algorithms on field programmable gate arrays (FPGA), Systems-on-Chip (SoC) and application-specific integrated circuits (ASIC), and unsupervised deep learning techniques for security assessment of hardware implementations. His research interests include security architecture of next-generation networks and security assessment and analysis of networks, devices and circuits using deep learning.
\subsection*{  } 
\noindent \textbf{Jithin Jagannath}  is the Chief Technology Scientist and the Founding Director of the Marconi-Rosenblatt AI/ML Innovation Lab at ANDRO Computational Solutions, LLC. He is also the Adjunct Assistant Professor in the Department of Electrical Engineering at the University at Buffalo, State University of New York. He received his M.S. degree in Electrical Engineering from University at Buffalo; and received his Ph.D. degree in Electrical Engineering from Northeastern University. Dr. Jagannath heads several of the ANDRO's research and development projects in the field of 5G and beyond, signal processing, MIMO beamforming, SIGINT, applied machine learning, protocol design for ad hoc networks, Internet-of-Things, and UAV automation for several customers including U.S. Army, Navy, DHS, and USSOCOM. He is currently leading several teams developing commercial products such as SPEARLink\texttrademark~ among others. Dr Jagannath was the recipient of 2021 IEEE Region 1 Technological Innovation Award with citation, "For innovative contributions in machine learning techniques for the wireless domain". He is also the recipient of AFCEA International Meritorious Rising Star Award for achievement in engineering and AFCEA 40 Under 40 in 2022.

Dr. Jithin Jagannath actively serves the academic community by providing 100+ invited technical paper reviews to several leading journals and conferences in the field of wireless communication, machine learning,  and signal processing.  He also serves as the Adjunct Assistant Professor, Department of Electrical Engineering at the University at Buffalo. He has been invited to give scholarly talks around the world including the latest Keynote titled,"Demystifying the Spectrum with Machine Learning for Beyond 5'' at an IEEE conference. Dr. Jagannath's recent research has led 10 Patents (granted, pending and provisional). Dr. Jagannath is an IEEE Senior member and serves as an IEEE SPS Applied Signal Processing Systems Technical Committee member. He has been invited to sever as TPC member of several leading conferences.  



\end{document}